**System parameters and measurement instrument parameters are not separately observable: Relational mass is observable while absolute mass is not**

Craig R. Holt*

Weaverville, California, USA

November, 2013

*ABSTRACT*

A brief summary of the objections to the relational nature of inertial mass, gravitational mass and electric charge is presented. The objections are refuted by showing that the measurement process of comparing an instrument reference clock and a reference rod both obeying the laws of physics to a system obeying the same laws of physics results in relational quantities: inertial mass, gravitational mass and electric charge appear only as ratios. This means that scaling of the absolute inertial mass of every object in the universe by the same factor is unobservable (likewise for gravitational mass and electric charge). It is shown that the measurement process does not separate the instrument parameters from the system parameters. Instead a measurement produces functions of fundamental, dimensionless parameters such as the fine structure constant, electron-proton mass ratio and the proton gyro-magnetic factor. It is shown that the measurement of Planck's constant also results in such a function of these dimensionless parameters. Application of this analysis to the "absolute" Planck units shows that they too are functions only of relational quantities. Analysis focuses initially on non-relativistic systems for the sake of clarity and simplicity. The results are then extended to relativistic systems described by the Dirac equation and the Einstein field equations of general relativity. The results support the assertion that instrument parameters cannot be separated from system parameters even in classical mechanics.

## 1.0 Introduction

The idea that relational quantities are observable while absolute quantities are not has been around in the physics literature for some time [1,2,3,4]. For instance, Laplace observed [3], "The simplicity of the laws of nature therefore only permits us to observe the relative dimensions of space," and Mach observed [4], "The ratio of the masses is the negative inverse ratio of the counter-accelerations." After Einstein's introduction of general covariance [5], it was generally accepted that absolute position and absolute motion are unobservable and that only relative positions and relative motions are observable. Furthermore, there are significant ongoing efforts to develop purely relational theories of classical particle dynamics, gravity and quantum mechanics [6,7,8,9,10,11,12 for example]. The idea of relational dynamics takes the form of 'background independence,' which is identified by some authors [13, 14] as a crucial requirement for the unification of quantum mechanics and general relativity.

Elsewhere in the literature, in spite of the observations of Leibnitz, Berkeley, Laplace and Mach, doubts remain regarding the purely relational status of duration, length, mass and charge. In fact, a number of challenging arguments can be advanced to make the opposite case for mass. The magnitude of the gravitational acceleration on the moon is different from that on earth, suggesting that absolute gravitational mass is observable [15,16]. The inertial mass can be measured without explicit comparison to a reference mass by measuring the de Broglie wavelength of a moving particle or by measuring the recoil velocity of an atom after photon absorption [17,18]. The concept of Planck units [19,20] seems to suggest that values of absolute mass, length and duration can be quantified without physical references. The key point of these objections is that a relational mass does not arise because none of them require two interacting masses as must occur in Mach's method of measuring the "counter-accelerations."

Another apparently convincing argument, introduced in this paper, against relational mass is based on the observation that different clocks vary in their dependence on inertial mass. The frequencies of macroscopic electrical clocks are inversely proportional to the square root of inertial mass, orbital frequencies of gravitating bodies are proportional to the square root of gravitational mass, atomic clock frequencies are proportional to inertial mass, and light clock frequencies don't depend on mass at all. Therefore, comparisons among clocks ought to show an observable dependence on absolute mass.

This paper addresses the doubts raised by these objections by analyzing the classical measurement process. The analysis incorporates physics-based models of clocks and rods that are to be compared to physical systems obeying the same laws of physics.

The argument against the objections is based on three key observations. First, one must be careful to distinguish between absolute quantities and relational quantities. Second, if one expresses the dynamical equations describing some system as functions of its absolute quantities, the measurement process necessarily involves a comparison between the system and an instrument described by the same dynamics [21]. Third, the dependence of both the instrument clock and the instrument rod (for length measurements) on absolute mass and absolute charge must be taken into account in modeling the measurement process. For instance, when an atom-based measuring rod is used to set the mirror separation in a light clock, the resulting clock oscillation frequency actually does depend on inertial mass because the length of the rod depends on the electron mass by virtue of quantum theory.

This paper shows that the measurement process quantifies relationships among systems and measuring instruments but does not quantify absolute parameters belonging separately to either. This is done by developing a simple model for actual instruments and then analyzing the result of measurements in order to ascertain the dependencies. The results of the analysis show that even classically there is a sense in which the instrument cannot be separated from the system it is being used to measure. Measurements simply do not provide quantification of absolute properties of the system independent from the absolute properties of the instruments. It will be shown that they provide relational values for mass, charge and length as well as measurements of the coupling constants which, as will be argued, are relational measures of interaction strength.

The methodology is applied to the measurement of the fundamental constants of nature such as Planck's constant, the speed of light and Planck units. A surprising (though not to some [19,22]) consequence of this classical instrument-system inseparability is that measurements of these fundamental constants necessarily depend on instrument parameters characterized by relational quantities such as coupling constants and the electron-proton mass ratio. The meaning of this result is subtler than the simple fact that the numerical values of fundamental constants depend on the choice of units (for example mks vs. cgs). More profoundly, the result argues against the view that Planck's constant, the speed of light, the gravitational constant and the constant of electromagnetism are fundamental constants (in agreement with Duff [23]) because they are absolutes that cannot be observed. Instead, it argues that only dimensionless, relational quantities are observable.

## 2.0 The objection to relational mass based on clock rate analysis

**2.1 The simple argument for relational quantities and against absolute quantities**
The equations of physics are typically displayed without explicitly showing or stating the physical units. When a mass variable, $m$, is used, for instance, it actually represents a relational value because it is assumed to have physical units; that is, $m = m' / m'_R$ where $m'$ is the absolute mass and $m'_R$ is a reference absolute mass that is assumed to exit somewhere. This distinction between relational and absolute quantities is the fundamental reason behind the assertion that all experimental measurements (observables) are relational and that absolute quantities are unobservable.

A measurement of length, for instance, does not produce two numbers—one for the absolute length of the system and one for the absolute length of the measuring rod. Though a length measurement can be represented symbolically by a ratio, $L = L' / L'_R$, this single number is not actually calculated by dividing two previously measured absolute quantities. The implication of the ratio, however, is that scaling all absolute lengths by a single number is unobservable, which in turn means that absolute quantities have no physical meaning. This is analogous to absolute position in space-time having no physical meaning.

**2.2 Objections to the simple argument**
Though this straightforward argument may be convincing, it contradicts common sense, which says absolute mass does have an impact on our observations. If one replaces an object oscillating at the end of a spring with something heavier, the new oscillation frequency is quantifiably smaller. This situation appears to be different from the length measurement because these masses do not interact with each other to produce a single measurement as required by Mach's definition. It seems we do have two measurements—the two oscillation frequencies—which can be turned into measurements of absolute inertial mass.

This objection is overcome by accounting for how the clock rate (oscillation frequency), depends on its own internal inertial absolute masses. Consider a clock made from a spring identical to that in the system of interest but with a different absolute inertial mass, $m'_R$. The measured oscillation frequency of the system-object is equal to the number of cycles it passes through in a single clock cycle. Given the usual dynamical equations of a harmonic oscillator applied to absolute masses, the measured oscillation frequency of the system object is $\omega = \sqrt{m'_R / m'}$. This is clearly invariant under scaling of all absolute inertial masses, which supports the original assertion that absolute inertial masses are unobservable.

The flaw of this argument, however, is that it assumes that all clock rates are inversely proportional to the square root of some internal absolute inertial mass. The frequencies of atomic clocks, light clocks and orbiting gravitating bodies have a completely different dependence on absolute mass. The measured oscillation frequency of the system-object using these alternate clocks is not predicted to be a ratio of absolute masses, which contradicts the assertion that scaling of absolute inertial masses is unobservable. This objection can, however, be overturned by accounting for the dependence of measuring rods on absolute inertial mass. This observation may appear to be irrelevant to clock comparisons, yet as is shown in Section 4.0, it is crucial in establishing the relational nature of inertial mass.

Suppose the equations of physics are valid when expressed in terms of absolute quantities instead of relational quantities as was assumed in Section 2.1. The question investigated in this paper is, "Are there any measurements that result in the quantification of absolute quantities (specifically absolute mass) or do all measurements necessarily result in relational quantities in spite of the basic equations depending only on absolute quantities?" The analysis of this issue requires clear notation to distinguish absolute quantities from relational quantities. Throughout this paper, a variable with a prime on it indicates that it is an absolute quantity, and a variable without a prime indicates that it is relational.

**3.0 Analyzing clock measurements using atomic clocks and atomic rods**

A measurement of a system frequency, $\omega'_S$, using a reference clock frequency, $\omega'_R$, is obtained by comparing their values. The measured frequency, $\omega_S$, takes the form of a ratio, $\omega_S = \omega'_S / \omega'_R$. Likewise a measurement, $L_S$, of a system length, $L'_S$, using a reference rod length, $L'_R$, is a ratio $L_S = L'_S / L'_R$. This section develops physics-based models for reference clocks, reference rods and system clocks to establish the parameter dependencies of these measured relational quantities.

**3.1 Model of atomic reference clock and atomic reference rod**

The dynamical equations of physics apply equally to systems and to instruments used to make measurements on systems. One must take this into account when analyzing clock comparisons. The rate at which instrument clocks tick and the length of rods used to measure length are expected to change if all the absolute inertial masses and absolute electric charges changed proportionally. A careful analysis of measurements based on atomic clocks and atomic rods made out of atoms, will show that in fact these absolute quantities are not observable.

Currently metrologists construct reference atomic clocks based on emissions in atoms (e.g., hyperfine structure in cesium-133) and define reference lengths in terms of the wavelength of the emitted photons [24]. Generally, the atomic clock angular frequency is modeled by [24,25]

$$\omega'_A = \eta \omega'_0 , \qquad (3.1)$$

where $\omega'_0$ is an absolute angular frequency defined by

$$\omega'_0 \equiv \frac{1}{2}\alpha^2 \frac{m'_e c'^2}{\hbar'} , \qquad (3.2)$$

and the fine structure constant is defined in the usual way $\alpha = K'e'^2 / \hbar'c'$, where $K'$ is the Coulomb constant (note that $\omega'_0$ is related to the absolute Rydberg wavenumber, $R'_\infty$, according to $\omega'_0 = 2\pi c' R'_\infty$). The parameter $\eta$ is a dimensionless quantity that depends on the two quantum states selected for the emission spectrum. It is not primed because it is dimensionless.

For example, the transition between the $2p_{1/2} \rightarrow 1s_{1/2}$ in the hydrogen atom [26] $\eta$ can be computed to second order in $\alpha$

$$\eta = \left(1 + \frac{m'_e}{m'_p}\right)\left(\frac{3}{4} + \frac{11}{128}\alpha^2\right) , \qquad (3.3)$$

where $m'_e / m'_p$ is the electron-to-proton mass ratio. Yet another example is if the two quantum levels correspond to hyperfine transitions in the ground state of the hydrogen atom,

$$\eta = \frac{16}{3} g_p \frac{m'_e}{m'_p} \alpha^2 ,  \qquad (3.4)$$

where $g_p$ is the gyro-magnetic factor for the proton. Generally, even for atoms other than hydrogen, $\eta$ depends on dimensionless numbers such as $\alpha$, $m'_e/m'_p$, $g_n$ (gyro-magnetic factor for the nucleus), quantum numbers, the number of protons and the number of electrons in the atom ([25] argues for this model of $\eta$ based on cesium-133 hyperfine emissions).

The frequency of the atomic clock photon defines a reference angular frequency, $\omega'_R$, and a reference time, $T'_R$. A reference absolute oscillation angular frequency is defined in terms of the emitted photon frequency,

$$\omega'_R \equiv \omega'_A ,  \qquad (3.5)$$

and an absolute reference duration, $T'_R$, is defined to be a set number of reference oscillation periods corresponding to the photon emission frequency,

$$T'_R = N_T \frac{2\pi}{\omega'_R} .  \qquad (3.6)$$

Metrologists defined the unit of 'one second' by setting $N_T = 9,192,631,770$ when the two energy levels in the atomic clock are selected based on the hyperfine structure of cesium-133 [25].

The standard reference length is defined by metrologists to be a specific number, $N_L$, of wavelengths of the emission photon [27],

$$L'_R \equiv N_L \frac{2\pi c'}{\omega'_A} .  \qquad (3.7)$$

This is referred to as the *atomic rod* in this paper. The value of $N_L$ is selected by defining 'one meter' to be such that the distance light travels in one "second" is equal to $299,792,458$ meters. The distance traveled in the reference time (one second) is $c'T'_R$ and the measurement of this distance using the atomic rod is a ratio that must satisfy $c'T'_R / L'_R = 299,792,458$. Substituting (3.6) and (3.7) into this equation results in $N_T / N_L = 299,792,458$ or $N_L = 30.6633190$. The choice of both $N_T$ and $N_L$ are chosen to be consistent with the measured speed of light based on earlier standards. It will be convenient later to define the relational speed of light by

$$c = \frac{N_T}{N_L} = 299,792,458 .  \qquad (3.8)$$

The units of m/s are intentionally left out to be consistent with the notation that any variable without a prime is relational.

The crucial feature of the model of the reference length is in how it depends on absolute quantities and how it depends on relational quantities. Combine (3.1), (3.2) and (3.7) to obtain

$$L'_R = 4\pi \frac{N_L}{\eta\alpha} a'_0 ,\qquad(3.9)$$

where the absolute Bohr radius is defined by

$$a'_0 = \frac{\hbar'}{m'_e c'}\frac{1}{\alpha} .\qquad(3.10)$$

This suggests that a simple model for the atomic rod reference length be given by

$$L'_R = \chi a'_0 ,\qquad(3.11)$$

where the dimensionless proportionality parameter is defined by

$$\chi \equiv 4\pi N_L / \eta\alpha .\qquad(3.12)$$

Explicitly substituting in the value for the Bohr radius,

$$L'_R = \frac{\chi}{\alpha}\frac{\hbar'}{m'_e c'} .\qquad(3.13)$$

This is sometimes referred to as the 'Bohr ruler' [28], but here it is referred to as an atomic rod in order to emphasize that it models an actual instrument as opposed to just a mathematical definition. Keep in mind that $\chi$, like $\eta$, depends only on relational quantities as well as the arbitrary metrology choice of $N_L$.

This modern atomic rod bears a crucial similarity to the earlier standard rod defined by a particular bar of Pt-Ir [19,22,28,24]. A reasonable model of this measuring rod is that it is comprised of atoms separated by some multiple of the Bohr radius [24]. This makes sense because in the time-independent Schroedinger equation one can normalize the spatial coordinates by the Bohr radius to obtain a relational Schroedinger equation. This suggests defining an absolute reference length that is proportional to the absolute Bohr radius similar to (3.11),

$$L'_R = \tilde{\chi} a'_0 .\qquad(3.14)$$

The main difference is that the model for $\tilde{\chi}$ has different dependencies than the model for $\chi$. The parameter $\tilde{\chi}$ can depend on different factors such as the quantum structure of solids, the number of atoms in the Pt-Ir bar, and environmental factors. However, both reference length models, (3.11) and (3.14), have the same dependence on absolute quantities, and $\tilde{\chi}$ depends on dimensionless quantities.

The absolute quantities in the atomic reference clock and atomic reference rod can be summarized in terms of the Compton wavelength defined by

$$\lambda'_C = 2\pi \frac{\hbar'}{m'_e c'} .\qquad(3.15)$$

The atomic reference angular frequency is obtained by combining (3.1),(3.2) and (3.5)

$$\omega'_R = \pi\eta\alpha^2 \frac{c'}{\lambda'_C} ,\qquad(3.16)$$

and the atomic reference rod is given by trading out the Bohr radius from (3.11) for the Compton wavelength

$$L'_R = \frac{1}{2\pi}\frac{\chi}{\alpha}\lambda'_C . \tag{3.17}$$

The relationship between the atomic rod and atomic frequency given by (3.12) results in an expression for the atomic reference rod length in terms of $\eta$,

$$L'_R = \frac{2N_L}{\eta\alpha^2}\lambda'_C . \tag{3.18}$$

The parameters $(\eta, \chi)$ are dimensionless parameters that depend on the choice of quantum levels in the atomic clock. These three equations represent the models of the reference clock and reference rod used throughout this paper to compare with system frequencies and system lengths.

It is important to distinguish between models for actual clocks and actual rods from the mathematical concept of 'atomic units'. They are similar in that the reference length (3.11) is proportional to the Bohr radius and the reference duration (3.6) is inversely proportional to $\omega'_0$, but they are crucially dissimilar in that the 'atomic units' do not necessarily model an actual existing instrument. If there were such an instrument, the measuring rod must have $\chi = 1$ and the clock must have $\eta = 4\pi N_T$. For such a clock and rod, their $(\eta, \chi)$ parameters would have to be independent of any physical parameter. This requirement is not met by any of the atomic clocks nor by any of the atomic rods discussed above. Though the concept of 'atomic units' may be a convenient mathematical construct, it is a mistake to use these units to model measurements obtained from actual instruments.

**3.2 Light-clocks, clocks based on an LC oscillator and inertial clocks**

The light-clock is comprised of two mirrors separated by an absolute distance, $L'_c / 2$. The absolute angular frequency of this clock is defined in terms of the oscillation frequency characterizing the two-way-transit time,

$$\omega'_c = 2\pi \frac{c'}{L'_c} . \tag{3.19}$$

The mirror separation is measured using the reference rod. The measured value is obtained by comparing the system length to the reference length (as discussed earlier),

$$L_c = \frac{L'_c}{L'_R} . \tag{3.20}$$

The choice of the number, $L_c$, is an arbitrary choice made by the light-clock designer. Replace $L'_c$ in (3.19) with its value in (3.20), and use the model for the atomic rod (see (3.13), (3.2)) to obtain

$$\omega'_c = \frac{4\pi}{\chi\alpha L_c}\omega'_0 . \tag{3.21}$$

In spite of what one may suspect, the light clock absolute angular frequency actually *does* depend on absolute inertial mass due to its dependence on the frequency, $\omega'_0$ (see (3.2)). This must occur because the light clock mirror separation is set using a rod that is dependent upon the quantum dynamics of atoms.

It can now be shown that a measurement resulting from the comparison of light clock oscillations with atomic clock oscillations will not, in fact, depend on absolute inertial mass. This relational measurement is a ratio of the frequencies in (3.21) and (3.1):

$$\frac{\omega'_c}{\omega'_R} = \frac{4\pi}{\chi \eta \alpha L_c} \quad . \tag{3.22}$$

None of the quantities appearing in Eq. (3.22) are absolute quantities because both $\chi$ and $\eta$ depend on the ratio of electron mass to proton mass, the fine structure constant and the proton gyro-magnetic factor. To emphasize this point, substitute (3.12) into (3.22) to obtain

$$\frac{\omega'_c}{\omega'_A} = \frac{1}{N_L L_c} \quad . \tag{3.23}$$

Therefore a comparison of a light clock with an atomic clock doesn't in fact depend on any physical parameters whatsoever. It depends only on arbitrary numbers set by the clock designers and metrologists, and it certainly does not depend on absolute inertial masses.

An analysis of another kind of clock constructed using a classical LC electrical oscillator can be quickly completed because it is conceptually identical to a light clock. To show this, consider a simple model for the inductor and the capacitor. The inductor is a cylindrical coil of absolute length $L'_{Ind}$, with absolute cross-sectional area, $A'_{Ind}$, and a total number of turns, $N$. The capacitor is two identical plates with absolute area, $A'_{Cap}$, separated by absolute distance $L'_{Cap}$ with no dielectric medium between. The absolute oscillation frequency of this LC circuit, given by $\omega'_{LC} = 1/\sqrt{LC}$, can be computed from the above model of the capacitor and inductor,

$$\omega'_{LC} = 2\pi \frac{c'}{L'_{LC}} \quad , \tag{3.24}$$

where the effective length scale for this LC circuit is defined by

$$L'_{LC} = N \sqrt{\frac{A'_{Ind} A'_{Cap}}{L'_{Ind} L'_{Cap}}} \quad . \tag{3.25}$$

The LC-clock is conceptually identical to the light clock as can be seen by comparing Eq. (3.24) with Eq. (3.19) and therefore it cannot be used to measure absolute inertial mass either.

The model of an inertial clock (both linear and rotational) is mathematically identical to a light clock and so the same conclusions follow. A linear inertial clock is a mass free of all external forces moving at constant velocity. Its absolute period is defined as the time it takes the object to traverse a fixed absolute distance, $L'_I$. This distance along its trajectory, which defines one period, is set to be some fixed number of reference rod lengths, similar to Eq.(3.20),

$$L_I \equiv \frac{L_I'}{L_R'} . \qquad (3.26)$$

This clock is conceptually identical to the light clock except the velocity of the mass is some fraction of the speed of light. The inertial clock velocity, $v_I'$, is set to be some fixed fraction of a reference speed, namely the speed of light; i.e.,

$$\beta_I \equiv \frac{v_I'}{c'} . \qquad (3.27)$$

This results in an absolute angular frequency for the inertial clock that is nearly identical in form to (3.19), namely

$$\omega_I' = 2\pi \frac{c'}{L_I'} \beta_I . \qquad (3.28)$$

Therefore, the result of comparing it to the atomic clock absolute frequency is identical to (3.23) except for the relational quantity $\beta_I$ and the use of $L_I$ instead of $L_c$

$$\frac{\omega_I'}{\omega_A'} = \frac{1}{N_L L_I} \beta_I . \qquad (3.29)$$

Once the inertial clock speed is arbitrarily set to be some fixed fraction, $\beta_I$, of the speed of light, scaling the absolute inertial mass is unobservable for the same reasons as in the light clock comparison.

A rotational inertial clock is mathematically identical (in this context) to a linear inertial clock. Suppose the rotational inertial clock is a rotating sphere (e.g., a planet). Setting the velocity of a point on the surface to be some fraction, $\beta_I$, of the speed of light, and setting the absolute circumference of the sphere to be some fixed number of absolute reference rod lengths, will also result in (3.29).

**3.3 Comparison of classical macroscopic electric clocks and gravitational clocks**
A macroscopic electric clock can be made, in principle, from two identical masses, $m_E'$ with equal charges, $Q'$, separated by a distance, $r_E'$, rotating about a common center. Its absolute angular frequency is given by

$$\omega_E' = \sqrt{2K' \frac{Q'^2}{m_E' r_E'^3}} . \qquad (3.30)$$

Radiation loss is ignored here in the spirit of keeping the clock comparison analysis simple. Similar to the light clock, the separation distance is measured using the reference atomic rod

$$r_E \equiv \frac{r_E'}{L_R'} . \qquad (3.31)$$

Use the reference rod defined by (3.11), and compare electric clock frequency to reference atomic clock frequency, (3.1),(3.5), to obtain

$$\frac{\omega_E'}{\omega_R'} = \sqrt{\frac{8}{r_E^3} \left(\frac{Q'}{e'}\right)^2 \left(\frac{m_e'}{m_E'}\right)} \sqrt{\frac{1}{\eta^2 \chi^3}} . \qquad (3.32)$$

All the quantities in (3.32) are relational, given that $\chi$ and $\eta$ do not depend on absolute quantities. Furthermore using (3.12) in this equation yields

$$\frac{\omega'_E}{\omega'_R} = \sqrt{\frac{1}{8\pi^3 N_L^3 r_E^3}} \sqrt{\left(\frac{Q'}{e'}\right)^2 \left(\frac{m'_e}{m'_E}\right)} \sqrt{\eta \alpha^3} \ . \tag{3.33}$$

These two equations show that a comparison of a macroscopic electric clock to an atomic clock could in principle allow one to measure the fine structure constant, but it does not result in absolute quantities being observable. This includes electric charge because charges appear only in ratios to other charges.

Comparison of the oscillations of a classical macroscopic gravitational clock to an atomic clock can be similarly analyzed. Consider a gravitational clock composed of two identical bodies with absolute inertial masses, $m'_G$, identical absolute gravitational masses, $M'_G$, separated by an absolute distance, $r'_G$, rotating about a common center. Note that absolute gravitational masses are symbolized by capital letters, $M'$, and absolute inertial masses are symbolized by lower case letters, $m'$ (the equivalence principle is incorporated later). The absolute angular frequency of the gravitational clock is given by

$$\omega'_G = \sqrt{2G' \frac{M'^2_G}{m'_G r'^3_G}} \ . \tag{3.34}$$

Measure the separation distance using a reference atomic rod,

$$r_G = \frac{r'_G}{L'_R} \ . \tag{3.35}$$

The comparison of the gravitational clock to an atomic clock must yield an expression similar to (3.32) except the coupling constant for gravitational interactions must be included,

$$\frac{\omega'_G}{\omega'_R} = \sqrt{\frac{8}{r_G^3} \left(\frac{M'_G}{M'_p}\right)^2 \left(\frac{m'_e}{m'_G}\right)} \sqrt{\frac{1}{\eta^2 \chi^3}} \sqrt{\frac{\alpha_G}{\alpha}} \ , \tag{3.36}$$

where the coupling constant corresponding to the strength of the gravitational field is defined in terms of the gravitational mass of a reference mass, $M'_R$ (e.g., the proton mass, $M'_R = M'_p$, or the mass of an object that defines the standard kilogram),

$$\alpha_G \equiv \frac{G' M'^2_R}{\hbar' c'} \ . \tag{3.37}$$

Using (3.12),

$$\frac{\omega'_G}{\omega'_R} = \sqrt{\frac{1}{8\pi^3 N_L^3 r_G^3} \left(\frac{M'_G}{M'_R}\right)^2 \left(\frac{m'_e}{m'_G}\right)} \sqrt{\eta \alpha^3} \sqrt{\frac{\alpha_G}{\alpha}} \ . \tag{3.38}$$

The measurement expressed in (3.36) and (3.38) is not sensitive to absolute quantities either. It does, however, depend on another fundamental dimensionless constant of nature, namely the gravitational strength as expressed by the gravitational coupling constant.

The objection presented in [15,16] that asserts that the absolute gravitational mass is observable can now be addressed. The objection is essentially that absolute gravitational mass is observable because the gravitational acceleration on the moon is different from that on earth. The weak point in the argument is that it neglects the fact that acceleration measurements depend on the choice of a clock. As is shown here, such measurements will depend only on the ratio of the gravitational masses of the body of interest to a reference gravitational mass (e.g., a proton). While it is true that the gravitational coupling constant is formally defined in terms of a reference mass, this coupling constant characterizes something more fundamental, namely the strength of gravitational interactions. This can be measured, in principle, between two protons and then be used as a fundamental constant of nature.

The concept of the 'strength of gravitational interactions', as characterized by $\alpha_G$, takes on meaning only when comparing gravitational clocks with other clocks whose dynamics are determined by non-gravitational forces. This is easily demonstrated by comparing the frequencies of two different gravitational clocks. This also holds true when only comparing electric clocks among each other. Equation (3.33) shows that the fine structure constant can only be obtained by comparing an electric to an atomic clock. This is interesting because the frequency of both clocks is determined by electromagnetic interactions. It is the quantum nature of atomic clocks and atomic rods that allows one to measure two coupling constants—gravity and electromagnetism.

Even though the coupling constants, $(\alpha, \alpha_G)$, have an expressed dependence on the proton absolute charge and a reference gravitational mass, their role in physics is relational. They take on meaning only when dynamics based on different kinds of forces are compared. Further discussions on this point follow below.

**3.4 Comparison among different atomic clocks**

Comparison among different atomic clocks does not result in measurements of absolute quantities either. Consider, for instance, the comparison of an atomic clock, $\omega'_{A1}$, based on transitions between principle quantum numbers ($\eta_1$ defined by (3.3)) with an atomic clock, $\omega'_{A2}$, based on hyperfine structure with $\eta_2$ by (3.4). Then

$$\frac{\omega'_{A2}}{\omega'_{A1}} = \frac{\eta_2}{\eta_1} = \frac{\frac{16}{3} g_p \frac{m'_e}{m'_p} \alpha^2}{\left(1 + \frac{m'_e}{m'_p}\right)\left(\frac{3}{4} + \frac{11}{128}\alpha^2\right)} . \qquad (3.39)$$

Clearly, these kinds of measurements do not depend on absolute quantities even though it may be extremely useful for high precision measurements of $\alpha$, $m'_e/m'_p$, and $g_p$. In practice the comparison is done among atomic and molecular emission spectra instead of actually constructing atomic clocks [22].

**3.5 Initial relational lengths are independent of absolute quantities**

The conclusions up to this point assume that all relational lengths, such as $L_c$ in (3.23), and the relational velocities of the inertial clocks do not depend on absolute quantities such as inertial mass. This assumption can be justified by considering the concept of another universe different from the current one in that all the absolute inertial masses are larger by the same scale factor. The issue of this paper is: Would any measurements performed in this other universe be different? A fair comparison requires that the other universe have the same initial conditions as the current universe, otherwise the measurements would establish a difference even before any dynamical effects could provide clues as to differences in absolute inertial masses. Therefore, a fair assessment requires that the *measured* initial conditions be identical in the two universes; i.e., independent of absolute quantities.

In particular, the clock initial conditions as specified by the relational length measurements, such as $L_c$ in (3.20), and initial relational velocity measurements, such as in the inertial clocks, $v = v'/c'$, must be exactly the same in both universes. If they weren't, then the measured size of the clocks in the new universe would be different even before they started ticking. This line of reasoning justifies the assumption that the relational lengths and relational velocities in the clocks analyzed above must be invariant with respect to scaling of absolute quantities.

**4.0 Examination of other kinds of measurements**

The clock comparison analysis showed that such measurements yield only relational quantities (with the possible exception of the coupling constants, which will be discussed below). This section examines other kinds of measurements that one could reasonably expect to yield a dependence on absolute quantities.

**4.1 Can absolute inertial mass be measured based on the de Broglie wavelength?**

Some metrologists propose to measure the de Broglie wavelength or frequency in order to characterize the inertial mass of an object [17,18]. The idea is to measure the de Broglie wavelength of a moving mass and in that way determine the inertial mass without having to explicitly compare it directly to another mass. The practical purpose is to avoid having to select a specific object to define a reference mass such as a one-kilogram Pt-Ir bar or the mass of some elementary particle. Since this appears to avoid relational masses, it suggests that the resulting mass measurement may be an absolute quantity [18]. This approach, however, does not result in values for absolute masses, for the same reasons that occurred in the clock comparisons: Clocks and measuring rods depend on inertial mass, which implicitly come into play during the measurement process.

The absolute mass, $m'$, of a body moving with absolute velocity, $v'$, is related to the absolute de Broglie wavelength, $\lambda'$, by

$$m' = \frac{\hbar'}{(\lambda'/2\pi)} \frac{\sqrt{1-(v'/c')^2}}{v'} \quad . \tag{4.1}$$

This equation alone is insufficient to specify the value of the absolute inertial mass because one can't observe the absolute de Broglie wavelength. Since the measured de Broglie wavelength requires a reference rod characterized by an absolute reference length, (4.1) can be re-expressed by judicially inserting a reference rod absolute length

$$m' = 2\pi \frac{\hbar'/c'L'_R}{(\lambda'/L'_R)} \frac{\sqrt{1-(v'/c')^2}}{v'/c'} \quad . \tag{4.2}$$

Use the atomic rod as defined by (3.13) to obtain

$$\frac{m'}{m'_e} = 2\pi \frac{\alpha}{\chi} \frac{1}{\lambda} \frac{\sqrt{1-\beta^2}}{\beta} \quad , \tag{4.3}$$

where the measured de Broglie wavelength is defined by $\lambda = \lambda'/L'_R$ and the measured particle velocity is given by $\beta = v'/c'$. The absolute electron mass finds its way into the measurement process because the absolute length of the atomic rod depends on it. It has already been established that $\chi$ does not depend on absolute inertial mass because of (3.12) so that (4.3) becomes

$$\frac{m'}{m'_e} = \frac{1}{2} \frac{\eta\alpha^2}{N_L} \frac{1}{\lambda} \frac{\sqrt{1-\beta^2}}{\beta} \quad . \tag{4.4}$$

The $\eta$ parameter depends only on relational quantities such as $\alpha$, $m'_e/m'_p$, $g_p$ (as discussed in Section 3.1), but certainly not on absolute inertial mass. Before seeing the above analysis it would seem reasonable to assert that the measurement of the de Broglie wavelength of a particle is independent of a reference mass. But this is not the case.

**4.2 Can any measurement in Newtonian gravity determine absolute mass?**

It is reasonable to think that absolute gravitational mass is observable because the magnitude of gravitational effects appears to depend on the absolute mass of the source of the gravitational field. This is asserted in [15,16] by noting that the gravitational acceleration on the Moon is observably different than on the Earth. The flaw in this assertion, however, is that it neglects the fact that the gravitational constant is a measured quantity—it must be calibrated using relational measurements.

Consider the Newtonian theory of gravity expressed in terms of absolute quantities (indicated by the primes):

$$m'_i \frac{d^2(\underline{r}'_i)}{dt'^2} = -G' \sum_k M'_i M'_k \frac{1}{\left|\underline{r}'_i - \underline{r}'_k\right|^2} \hat{r}_{ik} \quad . \tag{4.5}$$

These equations appear to show that the dynamics of gravitating bodies depend on absolute quantities and hence should be observable. This conclusion is premature because it neglects the fact that these equations needed to be validated experimentally, and that means taking measurements. By inserting the relevant reference lengths and reference times, equation (4.5) can be expressed in terms of relational measurements,

$$\frac{m'_i}{m'_R} \frac{d^2(\underline{r}'_i/L'_R)}{dt^2} = -G \sum_k \left(\frac{M'_i}{M'_R}\right)\left(\frac{M'_k}{M'_R}\right) \frac{L'^2_R}{\left|\underline{r}'_i - \underline{r}'_k\right|^2} \hat{r}_{ik} \quad , \tag{4.6}$$

with relational time defined by $t = t'/T'_R$ and provided

$$G = G'M_R'^2 \left(\frac{T_R'}{L_R'}\right)^2 \frac{1}{m_R' L_R'} . \tag{4.7}$$

The prime is not used on $G$ in anticipation of showing that it is relational. Use the atomic clock model of Section 3.1, to obtain

$$G = \alpha_G \left(\eta \alpha^2 \frac{N_T^2}{4\pi N_L^3} \frac{m_e'}{m_R'}\right). \tag{4.8}$$

This constant $G$ is relational with the possible exception of the gravitational fine structure constant (see later discussion).

Another way to see that the constant $G$ must be relational is to investigate how one would measure it using (4.6). A conceptually simple calibration system is a two-body system with one mass (e.g., a planet), $M_B'$, being much larger than the other mass, and a second mass being a very small sphere with mass $m'$. Drop the ball an initial distance $L'$ from the large mass and measure its instantaneous acceleration, $a$, due to the larger body pulling on it. The measured acceleration (relational) at an instant of time is related to the gravitational constant via (4.6),

$$a = -G\left(\frac{m_R'}{m'} \frac{M'}{M_R'}\right) \frac{M_B'}{M_R'} \left(\frac{L_R'}{L'}\right)^2 . \tag{4.9}$$

Use the equivalence principle in relational form,

$$\frac{M'}{M_R'} = \frac{m'}{m_R'} , \tag{4.10}$$

and solve for the measured value of $G$ to obtain

$$G = -a \frac{M_R'}{M_B'} \left(\frac{L'}{L_R'}\right)^2 . \tag{4.11}$$

This equation shows that the gravitational constant used in the Newtonian theory of gravity is purely relational. This means that the experimentally validated equations used in the Newtonian theory of gravity, (4.6), do not depend on absolute masses.

It is interesting to note that this value of $G$ is relational in the sense that arbitrary scaling of absolute masses, distances, and durations leaves $G$ invariant. This in turn means that one aspect of the theory of Newtonian gravity, (4.6), is relational in the sense that it does not depend on magnitudes of absolute quantities such as mass, length and duration. Note, however, according to [9,29,30], Newtonian theory is not fully relational due to an explicit assumption of the preexistence of absolute space and absolute time and these authors propose some remedies.

Clearly a similar analysis can be performed on the equations of classical electrodynamics. The same method discussed above results in

$$K = \frac{K' T_R'^2 Q_R'^2}{L_R'^3 m_R'} , \tag{4.12}$$

which becomes

$$K = \eta \alpha^3 \frac{N_T{}^2}{4\pi N_L{}^3} \frac{m'_e}{m'_R} , \qquad (4.13)$$

after application of the models for an atomic clock and an atomic rod where the absolute reference charge is the absolute charge on the electron, $Q'_R = e'$. The forms of the classical electro-dynamical equations are relational in that they are similar in form to (4.6). The observables are relational charges, relational masses and the coupling constant.

**4.3 Analysis of Einstein's theory of gravity—general relativity**

The Einstein field equations can also be transformed into an all-relational form. The standard form for the field equations is given by

$$G'_{\mu\nu} = \frac{8\pi G'}{c'^4} T'_{\mu\nu} . \qquad (4.14)$$

The appearance of the gravitational constant, $G'$, needs to be modified slightly to accommodate the transformation into relational form. In order to recover Newtonian dynamics in the weak field limit, i.e.,

$$m' \frac{d^2 r'}{dt'^2} = -M' \nabla' \phi' , \qquad (4.15)$$

where $\phi'$ is the gravitational potential proportional to $G'$, one must replace

$$G' \to \frac{M'}{m'} G' . \qquad (4.16)$$

Since the equivalence principle allows the replacement

$$\frac{M'}{m'} = \frac{M'_R}{m'_R} . \qquad (4.17)$$

The Einstein field equations can be adjusted slightly to take the form,

$$G'_{\mu\nu} = \frac{8\pi G' M'^2_R}{c'^4} \frac{T'_{\mu\nu}}{M'_R m'_R} . \qquad (4.18)$$

The reference masses in this equation are redistributed in order to make the gravitational coupling constant readily apparent.

Since the absolute coordinates, $x'^\mu$, are assumed to all be in units of absolute length, relational coordinates, $x^\mu$, are obtained from the coordinate transformation

$$x^\mu = \frac{x'^\mu}{L'_R} . \qquad (4.19)$$

Therefore, the absolute stress-energy tensor transforms according to

$$T'_{\mu\nu} \to L'^2_R T'_{\mu\nu} , \qquad (4.20)$$

and the resulting field equations become

$$G_{\mu\nu} = \frac{8\pi G' M'^2_R}{c'^4} \frac{T'_{\mu\nu}}{M'_R m'_R} L'^2_R . \qquad (4.21)$$

The absolute Einstein tensor is replaced with a relational tensor due to the coordinate transformation.

It can be shown that the right-hand side of (4.21) is purely relational using the illustrative case of a relativistic fluid. The stress-energy tensor is given by

$$T'^{\mu\nu} = (\rho' c'^2 + p')u^\mu u^\nu + p' g'^{\mu\nu} , \qquad (4.22)$$

where $\rho'$ is the matter density, $p'$ is the pressure and $u^\mu$ is the four-velocity vector of the fluid. The components of the four-velocity, $u^\mu$, do not contain a prime because they are relational due to their definition as a ratio of quantities with units of length; i.e., $u^\mu = dx^\mu / ds$. Substitute this stress-energy tensor into the right-hand side of (4.21) and rearrange terms

$$G' M_R'^2 \frac{1}{c'^4} \frac{T'_{\mu\nu}}{M_R' m_R'} L_R'^2 = \left\{ L_R'^3 \frac{\left[(\rho' c'^2 + p')u^\mu u^\nu + p' g'^{\mu\nu}\right]}{M_R' c'^2} \right\} \left\{ \frac{1}{m_R' c'^2} \frac{G' M_R'^2}{L_R'} \right\} . \qquad (4.23)$$

Examining the units of the terms inside the two sets of {} brackets reveals that they are purely relational.

Application of the clock model of Section 3.1 further substantiates the relational feature of Einstein's field equations. Define a relational density and pressure according to

$$\rho = \frac{1}{c^2} \frac{L_R'^3}{M_R'} \rho' , \qquad (4.24)$$

$$p = \frac{L_R'^3}{M_R' c'^2} p' . \qquad (4.25)$$

Note the distinction between the absolute speed of light, $c'$, and the measured speed of light, $c$, as defined by (3.8). Using the equations of Section 3.1 results in the Einstein field equations taking the relational form

$$G_{\mu\nu} = 8\pi \frac{\alpha_G \alpha}{\chi} \frac{m_e'}{m_R'} \left[ (\rho c^2 + p)u^\mu u^\nu + p g^{\mu\nu} \right] . \qquad (4.26)$$

Using the definition of $\chi$ from Section 3.1, the Einstein field equations become

$$G_{\mu\nu} = 2 \frac{\eta \alpha_G \alpha^2}{N_L} \frac{m_e'}{m_R'} \left[ (\rho c^2 + p)u^\mu u^\nu + p g^{\mu\nu} \right] . \qquad (4.27)$$

Equations (4.23), (4.26) and (4.27) are different forms of the same equation, and they all show that the Einstein field equation in relational coordinates contains only relational quantities.

The results of this section apply for a more general stress-energy tensor. For the relativistic fluid, the relational stress-energy tensor is given by

$$T_{\mu\nu} = (\rho c^2 + p)u^\mu u^\nu + p g^{\mu\nu} . \qquad (4.28)$$

Similar to the above analysis, the general relational stress-energy tensor after the relational coordinate transformation is defined by

$$T_{\mu\nu} = \frac{L_R'^3}{M_R' c'^2} T'_{\mu\nu} \qquad (4.29)$$

(analogous to the term in the first {} brackets of (4.23)). It is appropriate to normalize the absolute stress-energy tensor by the reference gravitational mass—instead of the inertial mass—because in this context it plays the role of a source of the gravitational field. The Einstein field equation then takes a form analogous to (4.27),

$$G_{\mu\nu} = 2\frac{\eta \alpha_G \alpha^2}{N_L} \frac{m'_e}{m'_R} T_{\mu\nu} \ . \qquad (4.30)$$

Furthermore the dynamical equations, $T^{\mu\nu}{}_{;\nu} = 0$, must likewise be relational because all quantities appearing in this equation are relational.

**4.4 Analysis of quantum field theory measurements—Compton scattering**

A measurement that holds some promise to contradict the analyses in the previous sections is the measurement of the Compton scattering cross section. This may prove interesting because a chargeless and massless photon is scattered by a charged, massive electron. Therefore it is plausible that cross-section measurements may produce values for the absolute inertial mass and absolute charge.

Consider a measurement of the scattering cross section in terms of the incident photon frequency, $\omega'_i$, and the scattered photon frequency $\omega'_s$. It is given to lowest order of quantum electrodynamics by the Klein-Nishina formula [31] (incident polarizations averaged and scattered polarizations summed)

$$\frac{d\sigma'}{d\Omega} = \frac{1}{2}\alpha^4 a_0'^2 F^2 \left( F + \frac{1}{F} - \sin^2\theta \right), \qquad (4.31)$$

where $\theta$ is the scattering angle and $F = \omega'_s / \omega'_i$. An equivalent form for $F$ is given by

$$F = \frac{1}{1 + (\hbar'\omega'_i / m'_e c'^2)(1 - \cos\theta)} \ . \qquad (4.32)$$

Equation (4.31) shows that the cross section depends on only one absolute quantity, the absolute Bohr radius, because $F$ is purely relational, being a ratio of frequencies. Likewise, the alternate form, (4.32), also depends only on the same absolute quantity because

$$\frac{\hbar'\omega'_i}{m'_e c'^2} = \frac{1}{2}\left(\frac{\omega'_i}{\omega'_R}\right)\eta\alpha^2 \qquad (4.33)$$

is relational.

The measured cross section, $\sigma$, however, must be expressed in terms of a reference length of the atomic rod,

$$\frac{d\sigma}{d\Omega} = \frac{1}{L_R'^2}\frac{d\sigma'}{d\Omega} \ . \qquad (4.34)$$

Note that any angle measurements embedded in $d\Omega$ are also relational because they involve trig-functions set equal to ratios of absolute lengths. Therefore the measured Compton scattering cross section is given by

$$\frac{d\sigma}{d\Omega} = \frac{1}{2}\left(\frac{\alpha^2}{\chi}\right)^2 F^2\left(F + \frac{1}{F} - \sin^2\theta\right), \tag{4.35}$$

and application of the atomic rod model, (3.12), yields

$$\frac{d\sigma}{d\Omega} = \frac{1}{2}\frac{(\eta\alpha^3)^2}{(4\pi N_L)^2} F^2\left(F + \frac{1}{F} - \sin^2\theta\right). \tag{4.36}$$

This measurement, like the clock comparisons, does not yield a value for absolute inertial mass nor absolute electric charge.

The electron mass does not explicitly appear because the atomic clock absolute length is proportional to the absolute Bohr radius. The absolute electron mass appears only in the ratio $m'_e / m'_p$ (see (3.4) for example). The observables resulting from the scattering are: 1) $\omega'_i / \omega'_R$ which is a relational frequency, 2) $\eta\alpha^3$ which is also relational because it resulted from the comparison of the macroscopic electric clock to the atomic clock (3.33), and 3) $\eta\alpha^2$ through (4.33). These measurements provide values for $\eta$ and $\alpha$, but none of the measurements depend on the absolute inertial mass nor the absolute electric charge.

**4.5 Analysis of Dirac equation to show it is relational**

The only absolute length scale appearing in the Dirac equation for the electron is the Compton wavelength, $\lambda'_C$, of the electron. Therefore, any physical quantity resulting from the Dirac equation will depend only on this absolute length scale. The instrument reference length, $L'_R$, is also proportional to the Compton wavelength because $L'_R = \chi a'_0$ and $a'_0 = \lambda'_C / \alpha$. Therefore measurements of any length quantity coming from the Dirac equation, such as cross sections, must be purely relational because the instrument Compton wavelength will cancel with that coming from the Dirac equation.

This conclusion can be made explicit by transforming the Dirac equation into a form that depends only on two physical parameters: $\lambda'_C$ and $\alpha$. The conventional form for the Dirac equation for the electron in absolute units is given by

$$\left[-i\gamma^\mu\left(\hbar'\frac{\partial}{\partial x'^\mu} + i\frac{e'}{c'}A'_\mu\right) + m'_e c'\right]\psi' = 0. \tag{4.37}$$

The field equations for the vector potential are given by,

$$F'_{\mu\nu} = \frac{\partial A'_\nu}{\partial x'_\mu} - \frac{\partial A'_\mu}{\partial x'_\nu} \tag{4.38}$$

$$\frac{\partial F'^{\mu\nu}}{\partial x'_\mu} = J'^\nu, \tag{4.39}$$

with the current density defined by

$$J'^\mu = e'\overline{\psi}'\gamma^\mu\psi', \tag{4.40}$$

and the gamma-matrices satisfying the usual anti-commutation relations

$$\{\gamma^\mu, \gamma^\nu\} = 2\eta^{\mu\nu} \qquad (4.41)$$

($\eta^{\mu\nu}$ the Minkowski metric). These equations come from the Lagrangian density,

$$L' = \bar{\psi}'\gamma^\mu\left(i\hbar'c'\frac{\partial}{\partial x'^\mu} - e'A'_\mu\right)\psi' - m'_e c'^2 \bar{\psi}'\psi' - \frac{1}{4}F'_{\mu\nu}F'^{\mu\nu} . \qquad (4.42)$$

An unconventional form for the Lagrangian that yields the same dynamics is

$$\tilde{L}' = \bar{\psi}'\gamma^\mu\left(i\frac{\partial}{\partial x'^\mu} - \frac{e'^2}{\hbar'c'}\tilde{A}'_\mu\right)\psi' - \frac{m'_e c'}{\hbar'}\bar{\psi}'\psi' - \frac{1}{4}\frac{e'^2}{\hbar'c'}\tilde{F}'_{\mu\nu}\tilde{F}'^{\mu\nu} . \qquad (4.43)$$

The "new" vector potential is related to the "old" one according to

$$A'_\mu = e'\tilde{A}'_\mu , \qquad (4.44)$$

and

$$\tilde{F}'_{\mu\nu} = \frac{\partial \tilde{A}'_\nu}{\partial x'_\mu} - \frac{\partial \tilde{A}'_\mu}{\partial x'_\nu} . \qquad (4.45)$$

The relevant absolute quantities are much more apparent in this form of the Lagrangian density when one expresses it in terms of the fine structure constant and the absolute electron Compton wavelength (see (3.15)),

$$\tilde{L}' = \bar{\psi}'\gamma^\mu\left(i\frac{\partial}{\partial x'^\mu} - \alpha\tilde{A}'_\mu\right)\psi' - \frac{1}{\lambda'_C}\bar{\psi}'\psi' - \frac{1}{4}\alpha\tilde{F}'_{\mu\nu}\tilde{F}'^{\mu\nu} , \qquad (4.46)$$

and the field equations become,

$$\left[-i\gamma^\mu\left(\frac{\partial}{\partial x'^\mu} + i\alpha\tilde{A}'_\mu\right) + \frac{1}{\lambda'_C}\right]\psi' = 0 \qquad (4.47)$$

$$\frac{\partial \tilde{F}'^{\mu\nu}}{\partial x'^\mu} = \bar{\psi}'\gamma^\nu\psi' . \qquad (4.48)$$

Note that these two equations depend on only two physical parameters: $\lambda'_C$ and $\alpha$.

The absolute Compton wavelength is the only absolute physical parameter in these equations, and it is a length. This means that any measurable length scale or time scale obtained from the solutions must be proportional to the Compton wavelength. Since the act of measurement is a comparison of the reference length, $L'_R$, to this absolute quantity, the measurement must be proportional to the ratio $\lambda_C = \lambda'_C / L'_R$. Inserting the form of the absolute reference length, (3.18), one obtains,

$$\frac{\lambda'_C}{L'_R} = \frac{\eta\alpha^2}{2N_L} , \qquad (4.49)$$

which is purely relational. A specific example of this was done in the previous section on the analysis of the cross section for Compton scattering. The difference here, however, is that the relational conclusion must hold to all orders of the perturbation expansion in $\alpha$.

The relational nature of the Dirac equation can also be established through the transformation to relational coordinates: $x^\mu = x'^\mu / L'_R$. Then the electromagnetic vector potential transforms according to

$$\tilde{A}'_\mu = \frac{1}{L'_R} \tilde{A}_\mu, \qquad (4.50)$$

and the Faraday tensor transforms according to

$$\tilde{F}'_{\mu\nu} = \frac{1}{L'^2_R} \tilde{F}_{\mu\nu}. \qquad (4.51)$$

Substitute (4.51) into (4.48) and raise the indices of $\tilde{F}'_{\mu\nu}$ using the Minkowski metric to obtain

$$\frac{1}{L'^3_R} \eta^{\mu\alpha} \eta^{\nu\beta} \frac{\partial \tilde{F}_{\alpha\beta}}{\partial x^\mu} = \bar{\psi}' \gamma^\nu \psi'. \qquad (4.52)$$

Define a relational Dirac field according to

$$\psi' = \frac{1}{L'^{3/2}_R} \psi, \qquad (4.53)$$

and the resulting field equations become,

$$\left[ -i\gamma^\mu \left( \frac{\partial}{\partial x^\mu} + i\alpha \tilde{A}_\mu \right) + \frac{L'_R}{\lambda'_C} \right] \psi = 0 \qquad (4.54)$$

$$\frac{\partial \tilde{F}^{\mu\nu}}{\partial x^\mu} = \bar{\psi} \gamma^\nu \psi, \qquad (4.55)$$

where the indices on the Faraday tensor are raised using the Minkowski metric. These forms of the field equations are purely relational and they depend only on the relational length, (4.49), and the non-dimensional fine structure constant. The gamma-matrices in this equation still obey the anti-commutation relations (4.41).

**4.6 Measurement of the speed of light and Planck's constant**

When one measures the speed of light or Planck's constant, what is actually observed? The issue of whether or not Planck's constant and the speed of light are fundamental constants of nature and what is meant by that label is somewhat controversial (see for example [19,23,32,33]). What is of interest in this paper, however, is what is observable at the level of theories considered in this paper—not whether or not the observables are fundamental in the sense discussed in the references.

Consider first the explicit measurement of the speed of light. This is a trivial result due to the current definition of the "second" and the "meter". When using atomic clocks and atomic rods, metrologists define it to be exactly $c = 299,792,458\,\text{m/s}$. It is instructive, however, to derive this result using the analysis techniques of this paper.

A photon travels an absolute distance, $L' = c'T'$ in an absolute time duration $T'$. The measured speed of light, $c$, is obtained by dividing two measurements: the first is the ratio of the absolute distance traveled to the reference length and the second is the ratio of the absolute travel time to the reference time,

$$c = \frac{L'/L'_R}{T'/T'_R} . \qquad (4.56)$$

Since, $L' = c'T'$

$$c = \frac{c'T'_R}{L'_R} . \qquad (4.57)$$

Given the definition of the absolute references in Section 3.1

$$c = 4\pi \frac{N_T}{\chi \eta \alpha} . \qquad (4.58)$$

Taking this one step further using (3.12),

$$c = \frac{N_T}{N_L} . \qquad (4.59)$$

A measurement of the speed of light using atomic instruments does not result in quantifying the absolute speed of light, $c'$. This result is of no surprise to metrologists who defined references so that the measured speed of light is exactly equal to (4.59). The current reference standard (discussed in Section 3.1) sets $N_T / N_L = 299,792,458$.

The important point of this paper, however is that (4.58) and (4.59) show that one cannot separate properties of light from measurement instruments (as characterized by $\eta$ and $\chi$) and the fine structure constant. Which term in (4.58) uniquely corresponds to a property of light and which term uniquely corresponds to the atomic clock or the atomic rod? The dependence of the measured speed of light on the atomic rod is especially obvious when the atomic rod is a Pt-Ir bar, (3.14). In this case, $\tilde{\chi}$ would be a very complicated function of the structure of the measuring instrument. The situation is vastly simplified when the atomic rod is defined as a fixed number of wavelengths of the photons in the atomic clock. Then the measured speed of light, (4.59) is simply an arbitrary number determined by the community of metrologists—it is not a fundamental property of nature.

**4.7 Measurement of Planck's constant**
A fundamental significance of Planck's constant is in its occurrence in the uncertainty principle. The uncertainty relation, for instance, in terms of uncertainty in absolute position and momentum is given by

$$\Delta x' \Delta p' \geq \frac{1}{2} \hbar' . \qquad (4.60)$$

Any measurements of position and momentum must be relational. The left hand side is expressed in terms of relational measurements by dividing both sides by the appropriate reference measurements of a product of length and momentum. The result is

$$\left(\frac{\Delta x'}{L'_R}\right)\left(\frac{\Delta p'}{m'_R(L'_R/T'_R)}\right) \geq \frac{1}{2} \frac{\hbar' T'_R}{m'_R L'^2_R} . \qquad (4.61)$$

Any experiment based on the particle-wave nature of particles would yield a measurement of the left hand side which in turn would yield a measurement of the quantity on the right hand side. Thus define the measurable Planck constant by

$$\hbar \equiv \frac{\hbar' T'_R}{m'_R L'^2_R} \,.\tag{4.62}$$

Using atomic clocks and atomic rods defined in Section 3.1 one obtains

$$\hbar = 4\pi \frac{m'_e}{m'_R} \frac{N_T}{\eta \chi^2} \,,\tag{4.63}$$

and using (3.12) results in

$$\hbar = \frac{1}{4\pi} \frac{m'_e}{m'_R} \frac{N_T}{N_L^2} \eta \alpha^2 \,.\tag{4.64}$$

A measurement of Planck's constant essentially results in quantities that characterize the interaction of measurement instruments, as modeled in Section 3.1, with the system that is being used to measure Planck's constant. This is especially apparent in (4.63) if a Pt-Ir bar is used as the atomic rod instead of the wavelength of the photons in the atomic clock. Equation (4.64) along with the clock comparison analysis makes it clear that the measurement of Planck's constant is similar to the measurement of the speed of light, in that neither can be separated from instrument parameters. Though this influence of the choice of instruments on measurements is already known [24], it still serves to support the main assertion of this paper, namely that measurements do not separate out absolute system parameters from absolute instrument parameters.

The results in (4.59) and (4.64) impact the issue of whether or not Planck's constant and the speed of light are fundamental constants of nature. These results support the assertion that they are not, because measurements of $\hbar$ and $c$ are equivalent to measurements of $\alpha$, $m'_e / m'_R$, and $g_p$ (through the dependence of $\eta$). This supports the position that these latter parameters are more fundamental than Planck's constant and the speed of light. That the measurable Planck constant depends on the choice of instrument parameters—not just in the trivial sense of using different words such as cm vs. meters—is of no surprise to the metrology community [21], but it is not a generally accepted view in other communities (see [23] for example).

**4.7 Analysis of Planck units as absolute quantities**
It is important to distinguish between two uses of the phrase 'absolute length.' In much of the literature, it can mean an absolute length *scale* (e.g., Planck length) such that no phenomena can be smaller than it, or it can mean an absolute velocity (speed of zero-mass particles) such that no velocities can exceed it. In this paper, however, 'absolute length' is analogous to absolute position in that both are unobservable. These meanings can be confused when considering the concept of absolute units that seemingly imply that absolute quantities are observable. The flaw in this view, however, is that these so-called absolute quantities are quantified in physical units, such as cm, and so they must in fact be relational. This point is clarified in this section by analyzing Planck units.

Concepts of absolute units, such as Planck units, do not change the conclusion that values of absolute quantities cannot be measured. One might think that expressing mass in terms of absolute units means that the absolute mass of objects can now be observed. The flaw in this logic is that the expressions for the absolute units depend on other absolute quantities, none of which can be observed. The absolute Planck mass, for instance, can be expressed in terms of absolute fundamental constants as

$$M'_{Planck} = \sqrt{\frac{h'c'}{G'}} \ . \tag{4.65}$$

Its value cannot be computed because none of the absolute values on the right hand side are observable, as previously demonstrated, so absolute inertial mass expressed in terms of the Planck mass likewise can't be quantified.

A relational expression that defines what is conventionally meant by the Planck mass is obtained by dividing (4.65) by a reference gravitational mass. The result is

$$\frac{M'_{Planck}}{M'_R} = \sqrt{\frac{1}{\alpha_G}} \ . \tag{4.66}$$

This measurable value is obtained by the clock comparisons measurements analyzed earlier. Such measurements do not, however, yield a numerical value for the absolute Planck mass, $M'_{Planck}$. Instead of a quantification of an absolute mass, (4.66) should be thought of as a theoretical prediction of what would result should the reference mass be compared to some phenomena characterized by the Planck scales of mass.

The quantification of an absolute Planck length is likewise impossible. The conventional expression for the absolute Planck length in terms of absolute quantities needs a slight modification to account for the difference in meaning between absolute inertial mass and absolute gravitational mass. The Planck length and Planck mass are related by

$$\frac{G'(M'_{Planck})^2}{L'_{Planck}} = m'_{Planck} c'^2 \ . \tag{4.67}$$

Using (4.65) in (4.67) results in the following expression for the absolute Planck length,

$$L'_{Planck} = \frac{1}{\mu'} \sqrt{\frac{G'\hbar'}{c'^3}} \ , \tag{4.68}$$

where an additional absolute parameter, $\mu'$, is introduced that relates the absolute gravitational mass to the absolute inertial mass,

$$\mu' = \frac{m'}{M'} \ . \tag{4.69}$$

The equivalence principle asserts that $\mu'$ is independent of either of the masses on the right hand side. Conventionally one sets $\mu' = 1$, but this is not necessary for the analysis in this paper.

Like the absolute Planck mass, the absolute Planck length is not observable due to its dependence on other unobservable absolute quantities. A relational Planck length, however, is quantifiable through the measurement process of comparing the absolute Planck length to the atomic rod reference length,

$$\frac{L'_{Planck}}{L'_R} = \frac{1}{\mu'} \sqrt{\frac{G'\hbar'}{L'^2_R c'^3}} \ . \tag{4.70}$$

Using the atomic rods defined earlier, one obtains

$$\frac{L'_{Planck}}{L'_R} = \left(\frac{m'_e}{m'_R}\right) \sqrt{\frac{\alpha_G \alpha^2}{\chi^2}} \ . \tag{4.71}$$

Using the atomic rod defined by (3.12),

$$\frac{L'_{Planck}}{L'_R} = \frac{1}{4\pi N_L} \left(\frac{m'_e}{m'_R}\right) \eta \alpha^2 \sqrt{\alpha_G} \ . \tag{4.72}$$

Note that the irrelevant $\mu'$ cancels out in the derivation of these equations. These equations are interpreted in the same way as the equation for the relational Planck mass in that it is a theoretical prediction of the relational size of some phenomena that may someday be observed using instrument clocks and instrument rods.

Superficially (4.72) is surprising because it seems to say that the Planck length, considered to be a quantum property of space-time, depends on the fine structure constant. The surprise arises from confusing the concept of *absolute* Planck length with the concept of *measured* Planck length. As has been demonstrated repeatedly, measured parameters intimately depend on one's choice of instruments, and not just in the trivial sense of the labeling of units (meters, centimeters, millimeters). The measurement inseparably incorporates the dynamical properties of the system and the instruments, which are characterized in the case of (4.72) by the parameters ($\eta, \alpha, \alpha_G, m'_e / m'_R$). Had one used different clocks and rods with dynamics dominated by say the strong forces, the measured Planck length would depend on the strong force coupling constants. It is simply a matter of correctly modeling what is the result of the measurement for a given set of instruments. Finally, since the absolute Planck length is unobservable, one cannot know what it depends on separate from the instrument used to measure phenomena on this extremely small scale.

### 5.0 Relational interpretation of coupling constants

Do the two coupling constants, $\alpha, \alpha_G$, quantify the absolute electron charge and absolute gravitational proton mass or do they quantify a relationship? First consider an argument for absolute quantities. Setting the absolute values of the $c', \hbar', G', K'$ equal to unity (which is allowed because their individual values are unobservable), results in $\alpha = e'^2$ and $\alpha_G = M'^2_R$. These expressions strongly suggest that that these two coupling constants ought to be identified with the absolute electric charge and absolute gravitational mass respectively. Since the coupling constants were demonstrated to be observable, the absolute electric charge and absolute gravitational mass are observable, which contradicts a crucial assertion of this paper.

These expressions are, however, misleading because by expressing them exclusively as functions of particle properties one ignores an important aspect of their role in physics: they characterize the strengths of the basic forces of nature. The fine structure constant of electromagnetism is a measure of the *interaction* strength between charged particles and photons. The value of the fine structure constant is observable only because there is an interaction between charged particles and electromagnetic fields. Atomic clock frequencies and atomic rod lengths result from the interaction of electrons with the electromagnetic field generated by other charged particles. This interaction is modeled by the fine structure constant, which becomes observable when comparing a macroscopic electric clock with an atomic clock, (3.33), or when comparing different atomic clocks or emission spectra, (3.39).

The classical scattering of light by an electron occurs because the electromagnetic field of the propagating wave interacts with the electron, causing it to accelerate, which in turn acts as a source that generates an electromagnetic field (the scattered wave). The strength of the interaction is quantified by the measured low-frequency ($\omega'_s = \omega'_i$) Thompson-scattering-cross section (given by Eq. (4.35) with $F = 1$). It depends only on the coupling constant and the atomic rod properties as modeled by $\chi$. Finally, quantum field theory describes interactions using Feynman diagram vertices, which characterize interactions in terms of a coupling constant. This particle-particle interaction is what gives the coupling constants meaning.

Another justification for the relational interpretation is that measurements of coupling constants are obtained through an interaction of the measuring instrument and the system. It has been shown in this paper that a crucial feature of such measurements is that they do not provide enough information to completely specify any of the absolute parameters in the instrument nor any of the absolute parameters in the system. A crucial example of this is Compton scattering. It was shown that the absolute charge on an electron could not be isolated in measurements of cross sections—only the fine structure constant, $\alpha$, and the reference instrument parameter, $\eta$, resulted from the interaction. These are universal parameters that cannot be uniquely assigned to the electron nor to the cesium clock. Like all relational measurements, neither can they be unambiguously assigned to the system or the instrument. There simply are not enough measurements to uniquely and completely specify all the absolute parameters in both the system and the instruments. Therefore, all parameters produced by measurements must be interpreted relationally.

## 6.0 Summary of the analysis of measurements

Analysis of clock comparisons measurements showed that only relational quantities are observable. The measurements depend on relational measurements of inertial mass, gravitational mass and electric charge. They also depend on the coupling constants, which are also relational in spite of their deceptive and essentially symbolic expression in terms of absolute electric charge and absolute gravitational mass. The measurements also depend on arbitrary parameters set by metrologists in defining references and on arbitrary choices made by clock builders in selecting the physical, yet relational, parameters for the clocks.

The absolute inertial mass dependency of the model atomic rods and atomic clocks arises from the fundamental de Broglie wavelength of quantum mechanics, which defines a relationship between inertial properties characterized by momentum and spatial-temporal properties,

$$m'v' = \frac{h'}{\lambda'} \quad \text{or} \quad m'v' = \frac{\hbar'}{c'}\omega' \ . \tag{6.1}$$

An absolute reference length based on the de Broglie wavelength must be inversely proportional to the absolute reference mass, and an absolute reference frequency based on the de Broglie frequency must be proportional to absolute reference inertial mass, which is exactly the mass dependencies in the length of atomic rods (3.11) and the frequencies of atomic clocks (3.1). The de Broglie relations replace Mach's ratio of accelerations as the defining notion of relational inertial mass.

Analysis of measurements of a relativistic scattering cross section, Planck's constant, the speed of light, $G$, $K$ and Planck units did not change these conclusions. A curious result of this analysis is that measurements of Planck's constant, the speed of light and the Planck scale cannot be separated from measurements of the fine structure constants, the electron-proton mass ratio and the proton gyro-magnetic factor ratio. This occurs because even classically measurements inseparably combine instrument dynamics with the system dynamics.

**7.0 Scaling of absolute quantities vs. scale invariance**

The assertion, "The equations of physics are invariant with respect to scaling of absolute quantities," is not the same assertion as, "The equations of physics are scale invariant." For instance, consider the Klein-Gordon equation

$$\partial^\mu \partial_\mu \psi + \frac{1}{L_m^2}\psi = 0 \ , \tag{7.1}$$

where all the quantities appearing in this equation are assumed to be relational (i.e., $x^\mu = x'^\mu / L_R'$ and $L_m = L_m' / L_R'$). The relational length $L_m$ is the Compton wavelength of some mass, $m$. The scaling of all absolute lengths by a factor $\gamma$ means that absolute coordinates scale as $x'^\mu \to \gamma x'^\mu$, and the absolute reference lengths scale as $L_R' \to \gamma L_R'$. The Compton wavelengths also must scale as $L_m' \to \gamma L_m'$ because they can be used, in principle, as instruments to measure length, conceptually identical to using atomic rods to measure length. Therefore, if the lengths of all measuring rods are scaled, then all Compton wavelengths should be scaled in exactly the same way.

Equation (7.1) remains invariant under this transformation of absolute lengths because of the relational character of its parameters. It simply says that if the absolute size of everything were to get smaller, including all absolute Compton wavelengths, then the equation would be invariant, and the scaling of all absolutes would be unobservable.

In contrast, the issue of 'scale invariance' focuses on the question of whether or not dynamics are invariant when one considers regions of space-time that are either smaller or larger relative to some fixed reference length. This re-focusing is modeled by the scaling of the relational coordinates according to $x^\mu \to \gamma x^\mu$. A dynamical equation is considered scale invariant if $\gamma^a \psi(\gamma x)$ is a solution ('$a$' is a constant) whenever $\psi(x)$ is a solution. It can be shown that the Klein-Gordon equation is not scale invariant given this definition. This occurs because the equation contains a relational scale, namely the Compton wavelength.

## 8.0 Summary

This paper establishes that absolute quantities of inertial mass, gravitational mass and electric charge are unobservable. Only relational measurements are observable because the measurement process is a comparison between an instrument and a system.

Often a direct comparison between like quantities is not explicit, such as inferring mass by measuring the de Broglie wavelength, but the fact that the instrument obeys the same dynamical equations as the system means an implicit comparison between like quantities must occur. This is demonstrated extensively in the paper by analyzing the comparison of instrument rods and instrument clocks in a variety of systems. In each instance, it is shown that only coupling constants and ratios of absolute quantities are observable, not the actual absolute quantities. The extensive analysis is first performed on non-relativistic systems to establish confidence in the basic concept. Then an analysis of the relativistic Dirac equation and generally relativistic Einstein's field equations results in the same conclusions.

An argument is then presented to support the interpretation of the coupling constants as being relational. Interpreting the fine structure of electro-magnetism, for instance, as the square of the electric charge is misleading, because $\alpha$ can be measured only during an interaction of an electron with a photon (virtual photons in the case of interactions between charged particles). The coupling constant in macroscopic gravitational interactions is important only when quantifying gravitational dynamics using clocks based on non-gravitational interactions.

This paper is a deeper interpretation of the meaning of what has always been known in physics, namely that all measurements must have physical units. "Physical units" means that something "physical" must be used to compare with the system. The choice of reference instruments is arbitrary, but they are often chosen to be appropriate to a specific area of investigation. To think this is a mere convenience instead of a fundamental feature of nature is a mistake. The empirical fact is that only instrument-system relationships are observable, while the absolute parameters of an independent system and an independent instrument are not observable.

Even measurements of what are considered fundamental constants, such as $\hbar, c, G, K$ are shown to depend on instrument parameters and not just in the trivial sense of specifying whether the units are grams or kilograms. Analysis of such measurements reveals that they depend on the structure of the atomic clocks and atomic rods used to make the measurements in terms of their dependencies on relational quantities such as the fine structure constant or gyro-magnetic factors. Even a measurement of the Planck mass and Planck length is predicted to depend on the structure of the instruments.

The fact that relational quantities are the only fundamental observables is not new to many metrologists [21], some high-energy theorists [23] and groups working on quantifying their time dependence [22]. It is also accepted by some (but not all) in these same communities that Planck's constant and the speed of light are not fundamental physical constants—only such things as coupling constants and mass ratios among elementary particles are fundamental. This paper clearly substantiates this latter view by explicitly deriving formulas showing how measurements of Planck's constant or of the speed of light using atomic clocks and atomic rods only provide measurements of universal parameters such as the fine structure constant, the electron-proton mass ratio and the proton gyro-magnetic factor.

The analysis of this paper is a criticism of viewing classical phenomena as involving objects with their own absolute properties having an existence independent of measurement and interaction. That these absolute quantities are unobservable gives them the same ontological status as absolute position. The absolute inertial mass of a particle has no more physical meaning than its absolute position. Writing a measurement result as a ratio of absolutes, for instance $m = m' / m'_R$, is purely symbolic because neither the numerator nor the denominator can be known, which means the ratio cannot be calculated from the individual terms.

It is acknowledged that it is useful in solving physics problems to imagine absolute, independent entities: space-time as a pre-existing container and objects as having their own properties independent of interactions. Measurements, however, do not support this view because none of these entities in one's imagination are observable. Just as space-time points are given meaning by events comprised of coincidences of dynamical variables [5], physical properties are given meaning only by multiple relational instrument-system coincidences. If absolute properties are unobservable, then entities, as defined by these properties are unobservable. To say in spite of these facts that independent entities exist, is to say that something that is unobservable exists.

There is a significant community that accepts the relational nature of phenomena and who insist that this principle be incorporated into the foundations of physics (see [13] as a representative position), so the general conclusions of this paper should be of no surprise to this community. Nonetheless, there are two primary purposes to this paper. The first is to substantiate the relational nature of phenomena by addressing in detail a number of specific objections to the assertion that absolute inertial mass, absolute gravitational mass and absolute electric charge are unobservable. The second purpose is to show to show that even classically there is a sense in which the measuring instrument cannot be separated from the system being measured, even when measuring the so-called fundamental constants.

* Deceased. (Paper posted to arXiv by Michael G. Raymer, Department of Physics, University of Oregon.)